\documentclass[nofootinbib,preprint,aps,amsmath,superscriptaddress,tightenlines]{revtex4}
\usepackage{graphicx}
\usepackage{bm}
\usepackage{epsfig}
\usepackage{amsmath,latexsym}

\def\OMIT#1{}

\newcommand{\nn}{\nonumber}

\newcommand{\bea}{\begin{eqnarray}}
\newcommand{\eea}{\end{eqnarray}}

\begin{document}

\begin{flushright}
hep-th/0208060\\
UCB-PTH-02/39\\
LBNL-51486\\
CMUHEP-02-12
\end{flushright}

\title{\phantom{x}\vspace{0.5cm}
Effective Field Theory and Unification in AdS Backgrounds}

\author{Walter D. Goldberger}
\address{Department of Physics, University of California,
Berkeley, CA 94720}
\address{Theoretical Physics Group, Lawrence Berkeley National
Laboratory, Berkeley, CA 94720}
\author{Ira Z. Rothstein}
\address{Department of Physics, Carnegie Mellon University,
Pittsburgh, PA 15213}

\begin{abstract}
\vspace{0.5cm} \setlength\baselineskip{18pt} 

This work is an extension of our previous work, hep-th/0204160, which showed how to systematically calculate the high energy evolution of gauge couplings in compact $\mbox{AdS}_5$ backgrounds.  We first directly compute the one-loop effects of massive charged scalar fields on the low energy couplings of a gauge theory propagating in the AdS background.  It is found that scalar bulk masses (which generically are of order the Planck scale) enter only logarithmically in the corrections to the tree-level gauge couplings.  As we pointed out previously, we show that the large logarithms that appear in the AdS one-loop calculation can be obtained within the confines of an effective field theory, by running the Planck brane correlator from a high UV  matching scale down to the TeV scale.  This result exactly reproduces our previous calculation, which was based on AdS/CFT duality.  We also calculate the effects of scalar fields satisfying non-trivial boundary conditions (relevant for orbifold breaking of bulk symmetries) on the running of gauge couplings.

\end{abstract}
\maketitle


\newpage


\section{Introduction}
The long lifetime of the proton and the apparent convergence of the gauge couplings at a large scale has long been seen as strong evidence for an energy desert between the weak scale and an
underlying UV scale. The drawback of this picture is the difficulty in generating and stabilizing the requisite hierarchy of scales.   A minimal supersymmetric model seems to fit into such a paradigm, as it naturally stabilizes the hierarchy and leads to an even more compelling convergence of the gauge couplings.  However, generating the hierarchy necessitates some additional physics input.

Whatever physics is responsible for stabilizing the hierarchy should, at least naively, remain weakly coupled if it is to preserve the predictions of grand unification. Thus it would seem that we should expect all the physics up to the UV scale to be well described by perturbation theory. Here we instead explore the possibility that new strong dynamics may arise at the TeV scale without spoiling the successful low energy gauge coupling relations implied by  grand unified theories (GUTs).  In particular, we are interested in making low energy coupling constant predictions for grand unified theories which are embedded in the Randall-Sundrum (RS) scenario~\cite{RS1}.  In this model the hierarchy is generated geometrically, via a 5D anti-deSitter (AdS) metric, and can be stabilized via a mechanism such as that of~\cite{gw}.  If one works in conformal coordinates the metric of the  $\mbox{AdS}_5$ spacetime is given by
\begin{equation}
\label{eq:metric} ds^2 ={1\over (kz)^2} \left(\eta_{\mu\nu} dx^\mu
dx^\nu - dz^2\right),
\end{equation} 
where $z$ parameterizes location in the bulk spacetime, bounded by a UV (or ``Planck'') brane at $z_{UV}=1/k$ and an IR (or ``TeV'')  brane at $z_{IR}=1/T$, and $x^\mu$ denotes the four-dimensional Poincare coordinates parallel to the boundaries of the space.  The AdS radius of curvature is given by $1/k$.  For applications to the hierarchy problem, we take all dimensionful parameters to scale as appropriate powers of the 5D Planck scale $M$.  The weak scale is generated by the position of the IR brane, $m_W\sim 1/z_{IR}=T$ (where we measure mass scales in terms of the unit Minkowski metric, $\eta_{\mu\nu}$). 

We will imagine that at least the Standard Model gauge fields propagate in the background of Eq.~(\ref{eq:metric}).  Then a low energy observer can only measure the effective coupling to the Kaluza-Klein (KK) zero mode of 5D gauge fields.  However, recall that in compactified AdS, the low-lying KK modes of bulk fields have masses of order the scale $T$.  Thus $T$ serves as an effective compactification scale, and for the same reasons as in a compactified flat space field theory, the zero mode gauge coupling observable will be rendered uncalculable by strong coupling effects near this scale (for instance, the effects of higher dimensional operators give rise to corrections that go roughly as $q^2/T^2$ ).

These remarks would then lead one to believe that it is impossible to make any sense of effective field theory (EFT) at scales above the KK scale $T$.  This seems to imply  gauge unification at a large scale, of order the curvature scale $k$, is not meaningful in a field theoretic context.  In reality, however, as we emphasized in a previous paper~\cite{US}, this is not the case.  In compactified AdS, the strong coupling scale beyond which EFT breaks down is observable dependent.  While zero mode observables cannot be reliably analyzed via local field theory at energies larger than $T$, a correlator whose external points are localized at a coordinate $z$ in the bulk is perturbative up to an energy scale roughly given by
\begin{equation}
p_4\sim {M\over kz},
\end{equation}
where $p_4$ is a 4D energy scale.  In particular, a correlator localized on the Planck brane is calculable up to momenta of order the scale $M$.  Because of this, it is possible to test unification at energies of order $k$, since, using perturbative field theory, one can directly compute gauge field correlators whose external points lie on the Planck brane. When the momentum
drops below the scale $T$, the energy resolution is insufficient to distinguish between the Planck brane correlator and the zero mode correlator, and therefore the two become equal at low energies.  It follows that one can use the Planck brane correlators to make low energy predictions based on high energy GUT symmetry.  By analyzing the Planck correlator of a 5D gauge field via the AdS/CFT correspondence~\cite{maldacena,AdSCFT} as it applies to the RS model~\cite{Gubser,APR,RZ}, we argued previously~\cite{US} that the momentum dependence of this observable is quantitatively similar to what one finds in a purely 4D gauge theory, leading to low energy predictions as in 4D GUTs.  The possibility that gauge couplings in AdS backgrounds evolve logarithmically was first raised in~\cite{pomarol}, and in~\cite{rschw} within the context of a solution to the hierarchy problem\footnote{We disagree with both the methodology and the quantitative results of the authors of~\cite{rschw}, however.}.  More recently, in reference~\cite{ads}, the zero mode correlator  including a bulk mass was calculated using a Pauli-Villars regulator  with results which are consistent with our previous work \cite{US}, as well as with our results below. Coupling constant evolution in an AdS background was also considered from the point of view of supergravity low energy effective lagrangians in~\cite{choi}.   

In~\cite{US} we proposed an EFT calculation of low energy gauge couplings based on integrating out fields that acquire masses due to GUT symmetry breaking at a scale of order  $k$.  In this paper, we describe our EFT approach in more detail.  However, we first present an alternative non-decoupling computation of the effects of massive fields  on the  low energy gauge couplings (which is a technically trivial extension of our previous zero mass calculation).  That is, we will directly compute the one-loop corrections to the couplings of gauge field zero modes at momenta less than $T$ due to massive charged bulk fields in a compactified $\mbox{AdS}_5$ background (this has also been done using a variety of methods in~\cite{pomarol,rschw,choi,ads}).  In this calculation, the information about the UV physics (for instance physics around the GUT scale) is encoded in the dependence of the 4D effective gauge couplings on the bulk mass parameters of the heavy fields. In Section~\ref{sec:flat} we show how such a calculation is carried out in compactified 5D flat models.  For simplicity we will phrase much of our discussion in the context of a simple model with two $U(1)$ gauge fields and two charged matter fields.  In this toy example, a $Z_2$ symmetry that exchanges the gauge fields plays the role of GUT symmetry.  By breaking $Z_2$ either through non-symmetric scalar masses or by boundary conditions (in the case of field theories propagating on the flat 5D orbifold $R^4\times S^1/Z_2$), we obtain definite predictions for the relation between the low energy gauge couplings of the two $U(1)$ gauge fields.

The results of the analogous AdS computation are discussed in Section~\ref{sec:adszm}.  In contrast to flat 5D models, where the dependence of the low energy coupling on the heavy bulk masses is linear, in AdS low energy gauge observables are only logarithmic functions of 5D mass scales, as long as they are smaller than the curvature scale $k$ (this result was also obtained in~\cite{ads} using a Pauli-Villars regularization procedure).  Because the coefficient of the logarithm is the same as in a purely 4D theory, low energy predictions in our $Z_2$ model where symmetry is broken by bulk scalar masses are therefore the same as in the flat 4D version of the model.  We also discuss the case in which the $Z_2$ symmetry is broken by non-trivial boundary conditions.  When the symmetry is broken explicitly on the Planck brane only, large logarithms of the ratio of $\sqrt{q^2}$ to the curvature scale appear in low energy predictions.  No large logarithms appear when the source of symmetry breaking is the TeV brane.  

Although formulas such as those presented in Section~\ref{sec:adszm} are sufficient for working out low energy predictions in warped GUTs, the physical origin of the large logarithms is somewhat obscure.  Furthermore one might worry that because of the large one-loop corrections, ordinary perturbation is invalid for computing such quantities.  For perturbation theory to be sensible, there must be a way of employing the renormalization group (RG) to resum the large logarithms into scale dependent couplings.  In Section~\ref{sec:adsplanck}  we explicitly show how to calculate the two-point Planck gauge correlator for momenta larger than the KK scale.  We use this to describe how the results presented in Section~\ref{sec:adszm} may be understood from an EFT perspective in terms of a procedure in which massive bulk fields are integrated out when the external momenta in the correlators become of order their masses.  In this approach, the large logarithms arise due to running of effective couplings.  This result can also be obtained by employing AdS/CFT ideas, along the lines that we used in~\cite{US}.  A discussion of such an approach can be found in Section~\ref{sec:adscft}.

\section{Coupling Constant Evolution in Flat Space Models}
\label{sec:flat}

Before discussing gauge theories in $\mbox{AdS}_5$ backgrounds, it is useful to review the situation in flat 5D spaces of the form $R^4\times S^1$ or $R^4\times S^1/Z_2$.   In this paper we will take as our ``GUT'' model a $U(1)_1\times U(1)_2$ gauge theory with scalar matter fields $\Phi_i$ ($i=1,2$) charged under $U(1)_i$.  In this model, GUT symmetry is a $Z_2$ symmetry that interchanges the two scalars as well as the two gauge fields, and ensures the equality of the two $U(1)$ couplings in the UV. By breaking $Z_2$ either softly (through explicit symmetry breaking scalar mass terms) or through boundary conditions (in the case of theories on $R^4\times S^1/Z_2$) we obtain different predictions for how relations of the low energy effective couplings constants (defined in a suitable way; see below) deviate from $Z_2$ symmetric values.  While simplistic, this $Z_2$ model captures the relevant field theoretic aspects in a clear way.  

For the sake of completeness, let us first recall the standard calculation of the low energy coupling constant relations in four-dimensional theories using EFT~\cite{Weinberg}. Typically when calculating the evolution of couplings it is most convenient to work within the framework of an  EFT. By integrating out one scale at a time the calculation is greatly simplified, and the evolution is accomplished by sequentially running and matching across mass thresholds.   Working in an EFT also allows the implementation of mass independent subtraction schemes (e.g. $\overline{MS}$) which leads to a significant simplification of the solutions of the RG equations. For instance, suppose that in our toy model $Z_2$ is softly broken by the introduction of hierarchical scalar masses $m_1\gg m_2$.  Then for high energies, the $Z_2$ ``GUT'' symmetry will be manifest.  To derive relations between low energy ($E\ll m_2$) couplings to leading logarithmic (LL) approximation within an EFT framework, we first integrate out the heavy field $\Phi_1$ at a scale $\mu\sim m_1$.  To LL accuracy, this is accomplished by simply removing the field $\Phi_1$ from the theory, and setting the $U(1)$ couplings  to a common value  at the scale $m_1$.  We then run the couplings, by integrating the one-loop RG equations derived in the effective theory below the scale $m_1$.  Finally, below the scale $m_2$ we remove $\Phi_2$ from the theory, at which point the running of the couplings is completely frozen.  Following this procedure, one finds the standard result
\begin{equation}
\label{eq:eft4d}
{1\over g_1^2(\mu)}-{1\over g_2^2(\mu)} = {1\over 24\pi^2}\ln{m_2\over m_1},
\end{equation}
valid at LL order for all $\mu<m_2.$

Alternatively, in the original field theory, one could simply calculate the low energy observable of interest, keeping the full particle content in the graphs.  If one works in a mass dependent
scheme such as momentum subtraction, the decoupling of heavy modes occurs automatically, and there is no need to work within an EFT.  However, it is also possible to calculate in the full theory using a mass independent scheme.  In this case decoupling will not be manifest.  Nevertheless, because all physical predictions are independent of the specific scheme chosen to carry out the calculations, one can be assured of getting the correct result for low energy quantities. 

Recall how this works in the $\overline{MS}$ scheme.   In $\overline{MS}$, the RG equations do not properly account for the decoupling of heavy fields, leading to spurious RG evolution of the couplings at low energies.  This unphysical running of the couplings is compensated by logarithms of the heavy particle masses in low energy matrix elements.  The two effects exactly conspire to properly decouple the heavy modes from low energy physics.  As an explicit example of this consider again the $U(1)\times U(1)$ model.  If we define an effective gauge coupling in terms of the 1PI gauge field two-point correlator (doing so gives rise to a resummation of all leading one-loop logarithms from the 1PI graphs), we find, working in Euclidean signature
\begin{equation}
{1\over g_i^2(q^2)}= {1\over g^2_i(\mu)} -{1\over 16\pi^2}
\int_0^1 dx (1-2x)^2 \ln\left[{m_i^2+x(1-x)q^2\over\mu^2}\right],
\end{equation}
where in $\overline{MS}$
\begin{equation}
\mu {d\over d\mu} g_i(\mu) ={g_i^3\over 48\pi^2}.
\end{equation}
We see that the running couplings $g_i(\mu)$ do not properly incorporate the decoupling of the charged fields at the thresholds $m_1$, $m_2$.  However, the prediction for the difference in the gauge forces at
$q^2\ll m_1^2$ (as measured, say by a Wilson loop observable) is the same as in the EFT approach
\begin{eqnarray}
 {1\over g^2_1(q^2)} - {1\over g^2_2(q^2)} &=& {1\over 16\pi^2} \int_0^1 dx (1-2x)^2 \ln\left[{m_2^2+x(1-x)q^2\over m_1^2+x(1-x)q^2}\right]\\
\nn
 &=& {1\over 24\pi^2}\ln{m_2\over m_1} +{\cal O}(q^2/m_2^2),
\end{eqnarray}
where  we used that $g_1(\mu)=g_2(\mu)$ due to the $Z_2$ symmetry.  The logarithm of $m_1/m_2$ in this case is not from the running of the gauge couplings, but rather from the infrared behavior of the 1PI correlator.  Yet this result is equivalent to the EFT prediction of Eq.~(\ref{eq:eft4d}). In this paper, we will use dimensional regularization plus $\overline{MS}$ throughout.  We will calculate low energy predictions both in an EFT framework (in Section~\ref{sec:adsplanck}), and in a non-decoupling scheme in Section~\ref{sec:adszm}.

\subsection{Bulk Symmetry Breaking in 5D Models}

Now consider our toy model propagating in a flat, five dimensional (Euclidean) space compactified either on a circle of radius $R$ or on a line interval $S^1/Z_2$ of length $\pi R$.  We are interested in computing how five dimensional physics manifests itself at energies smaller than the compactification scale.  At such scales only the lowest KK eigenmodes of bulk fields are accessible as asymptotic states in scattering processes.  Consequently, the relevant observables are derived from the correlators of bulk field zero modes.  In particular, in a higher dimensional GUT theory we are interested in deriving relations between the gauge couplings as measured by a low energy observers.  Such gauge couplings can be defined in a number of ways, for instance as the function that multiplies the explicit ${1/{\vec q}^2}$ dependence of a Wilson loop whose perimeter is longer than the compactification radius.  Equivalently, one could define an effective $q^2$-dependent coupling in terms of the coefficient of the quadratic term in the 1PI action for the zero mode gauge field:
\begin{equation}
\label{eq:effg} 
{1\over g^2(q^2)} = {1\over g^2} + \Pi(q^2),
\end{equation}
 where $g$ is the tree-level coupling, and $\Pi(q^2)$ encodes one-loop effects (we show how to compute it by summing over the KK modes of bulk fields in Appendix~\ref{app:seff}).  Throughout most of our discussion, we will use the latter quantity as our definition of the coupling
constant measured by low energy observers.  Working in dimensional regularization, the one-loop vacuum polarization amplitude due to a charged 5D scalar (mass $m$) on a flat space of the form $R^4\times S^1$ is given by (see Appendix~\ref{app:seff})
\begin{equation}
\label{eq:s1}\Pi_{S^1}(q^2;R) = -{1\over 8\pi^2}\int_0^1 dx x\sqrt{1-x^2}\ln\left[2\sinh\left(\pi R\sqrt{{q^2 x^2/4}+m^2}\right)\right].
\end{equation}
which turns out to be finite.  Because the 5D gauge-coupling is dimensionful, $[g_5]=-1/2$, no logarithmic divergences are possible at one-loop.  Rather, divergences must scale linearly with an ultraviolet cutoff.  However, such divergences are simply set to zero in dimensional regularization, leading to our finite result.  While there are no UV logarithms, in the massless case finite logarithmic dependence can arise in the IR for $q^2 R^2\ll 1$
\begin{equation}
\label{eq:s1lim}\Pi_{S^1}(q^2;R)\simeq -{1\over 24\pi^2}\left[\ln\left(2 \pi R\sqrt{q^2}\right)-{4\over 3}\right] + \mbox{analytic in $q^2 R^2$}.
\end{equation}
Besides the logarithm of $\sqrt{q^2}$ all other momentum dependence for $\sqrt{q^2} R\ll 1$ is analytic in $q^2$.  This is exactly what we expect in the low energy limit, where our 5D theory should be describable by a 4D theory that consists of a massless scalar and a $U(1)$ gauge field.  In the four-dimensional limit $R\rightarrow 0$, the logarithm in Eq.~(\ref{eq:s1lim}) becomes singular, reproducing the usual one-loop UV logarithm of a 4D gauge theory.  All other $q^2$ dependence is analytic, which in a 4D EFT is encoded by the contribution of an infinite set of local operators involving the massless 4D modes.  It is also instructive to consider the limit $\sqrt{q^2} R\gg 1$ of Eq.~(\ref{eq:s1}).  In this case, the behavior of $\Pi_{S^1}(q^2;R)$ is dominated by a non-analytic power law term
\begin{equation}
\label{eq:pow}
\Pi_{S^1}(q^2;R)\simeq - {\sqrt{q^2} R\over 256}.
\end{equation}
Thus for large enough $q^2$, the one-loop correction to the coupling has the same size as the tree level result.  At this scale the effects of higher dimension operators are also unsuppressed.  This indicates that the 5D gauge theory becomes strongly coupled in the UV, and is no longer an adequate description of the physics. When $m\neq 0$, the low energy behavior of Eq.~(\ref{eq:s1}) depends on the size of $m$ relative to the compactification radius.  For $q^2<m^2$, and $m R <1$, we find
\begin{equation}
\Pi_{S^1}(q^2;R)\simeq -{1\over 24\pi^2}\ln \left(2\pi R m\right).
\end{equation}
Imagine that in our toy $U(1)\times U(1)$, $\Phi_1$ acquires a mass $m_1\ll R$ while $\Phi_2$ remains massless.  Then the deviations from $Z_2$ unification at low energy are found to be
\begin{equation}
\label{eq:mlR}
{1\over g_1^2(q^2)}-{1\over g_2^2(q^2)} \simeq{1\over 24\pi^2}\ln\left({\sqrt{q^2}\over m_1}\right).
\end{equation}
This is of course, what we expect to find if we decouple the tower of KK states by hand of and compute the relative running of the gauge couplings in a 4D EFT that includes only the lowest lying modes of the bulk fields.  Although our computation of $\Pi_{S^1}(q^2;R)$ included the contribution of all the KK excitations of bulk fields, we see that physics above the scale $m_1$ effectively decouples from physical quantities.  Then the apparent "running" is due simply to the mismatch between the zero mode and the lowest mode of the massive field. Above the lowest mode, the KK spectra of $\Phi_{1,2}$ essentially match. On the other hand if $m R\gg 1$, the $q^2 < m^2$ limit is given by
\begin{equation}
\label{eq:mgR}
\Pi_{S^1}(q^2) \simeq -{m R\over 24\pi} + {\cal O}(q^2/m^2).
\end{equation}
Then in the $Z_2$ model with $m_1 R\gg 1$ and $m_2=0$, one predicts for $q^2 R^2<1$
\begin{equation}
\label{eq:pred}
{1\over g_1^2(q^2)} - {1\over g_2^2(q^2)} \simeq -{m_1 R\over 24\pi},
\end{equation}
where we have ignored the logarithm of $\sqrt{q^2} R$ relative to the power correction proportional to $m_1 R$.  This result is directly applicable to models with TeV compactification scales, where unification of the $SU(3)\times SU(2)\times U(1)$ couplings at a low scale is achieved through power law ``running'' of the couplings~\cite{DDG}.  From the point of view of Eq.~(\ref{eq:mgR}), this running is not due to the RG evolution of the couplings (indeed for theories compactified on a circle, there are no logarithmic divergences and therefore in a mass independent scheme no RG flows to speak of).  Rather it comes about through finite, calculable corrections to the low energy couplings  that depend on a physical UV scale, in this case the bulk mass of heavy field.  At least in flat space models, the extreme UV sensitivity of Eq.~(\ref{eq:mgR}) implies that in general, low energy predictions are highly dependent on the exact nature of symmetry breaking near the UV scale.  Unknown UV physics can give rise to corrections that are as large as what appears on the RHS of Eq.~(\ref{eq:pred}).  For instance, if in our model the symmetry breaking mass $m_1$ arises dynamically, there are tree-level operators involving the $U(1)_1$ field strength whose contribution is of the same order as the quantum effects.  Eq.~(\ref{eq:pred}) must therefore be viewed as at best an order of magnitude estimate of the relation between low energy couplings, rather than a definite prediction for 4D physics based on symmetries of the 5D theory.

\subsection{Symmetry Breaking on Manifolds with Boundary}

Let us now discuss how the quantum corrections are modified due to the  inclusion  of  boundaries.  To be definite we will consider $U(1)$ gauge theories with scalar matter propagating on the orbifold $S^1/Z_2$, where the orbifold fixed planes (``branes'') bound the space.   On such space the short distance structure of the theory is modified because it is now possible to write down operators involving bulk fields that are localized on the boundaries.  Generically, we expect loop effects to generate new divergences that renormalize the coefficients of such operators.  Depending on the mass dimensions of the brane localized couplings, short distance divergences on the branes may be logarithmic, inducing non-trivial boundary RG flows~\cite{georgi}.

It is also possible to use the branes to break symmetries of the bulk theory.  This occurs because one may impose boundary conditions that are incompatible with the symmetries of the higher-dimensional lagrangian.  For instance, we may give fields within a given multiplet different boundary conditions.  Then from the 4D point of view there will be mass splittings among the low lying KK modes of the fields in the multiplet, so that a 4D observer will identify the compactification scale with the symmetry breaking scale.  On the other hand, the splitting among heavy KK multiplets (masses larger than the compactification scale) will be small relative to the mass, so that the symmetry is manifest at high energies.  In the 5D language this is just the statement that the symmetry is explicitly realized locally in the bulk, but broken at exceptional points.   

To illustrate these features, we consider the one-loop correction to the zero mode effective gauge coupling due to a scalar field $\Phi$ satisfying various types of boundary conditions on flat $R^4\times S^1/Z_2$ (we assume that the zero mode survives whatever boundary conditions we impose of the 5D gauge field).  First choose $(+,+)$ boundary conditions at the $Z_2$ points (located at $z=0,\pi R$, with $z$ a coordinate along the
$S^1/Z_2$)
\begin{equation}
\label{eq:ppbc}
\left.\partial_z\Phi(x,z)\right|_{z=0}=\left.\partial_z\Phi(x,z)\right|_{z=\pi R}=0.
\end{equation}
The one-loop vacuum polarization of the zero mode gauge field is
now given by (with $\epsilon=4-D$ and ${\bar\epsilon}^{-1}={\epsilon}^{-1}+{1\over
2}(-\gamma+\ln(4\pi))$
\begin{eqnarray} 
\nonumber 
\label{eq:ee}
\Pi^{++}_{S^1/Z_2}(q^2;R) &=& {1\over 48\pi^2\bar{\epsilon}} - {1\over 32\pi^2}\int_0^1 dx x\sqrt{1-x^2}\ln\left[{{q^2 x^2/4} + m^2\over\mu^2}\right]\\
& & {}+ {1\over 2}\Pi_{S^1}(q^2;R),
\end{eqnarray}
This result can be easily derived by noting that the lowest lying scalar KK state on the orbifold is in one-to-one correspondence with that on the circle, while there are half as many massive KK states on $S^1/Z_2$ as there are on $S^1$ (see Appendix~\ref{app:seff}).  Comparing to Eq.~(\ref{eq:s1}), we see that the orbifold calculation generates $1/\epsilon$ poles.  There
are now logarithmic divergences which renormalize new brane-localized gauge kinetic terms
\begin{equation}
\label{eq:locff}
{\cal L}_{5D} = {1\over 4}\left[\lambda_0\delta(z) +\lambda_R \delta(z-\pi R)\right] F_{\mu\nu} F^{\mu\nu}+\cdots,
\end{equation}
(note that given our normalization for the 5D gauge field, $\lambda_{0,R}$ is dimensionless). Loop effects therefore induce non-trivial RG flows for the boundary couplings.  It can be shown that \begin{equation}
\mu{d\over d\mu}\lambda^{++}_{0,R} = -{1\over 96\pi^2}. 
\end{equation}

We note that by simple power counting, the RG equations for the boundary couplings are saturated at one-loop\footnote{This is no longer true in more general models.  For instance, a bulk fermion with mass $m$ can give rise to a two-loop correction to the beta function proportional to $m g_5^2$.  Brane localized fields coupled to the bulk gauge field may also contribute to the RG equations starting at one-loop.}.  One can similarly calculate the one-loop effects of $(+,-)$ scalars
\begin{equation}
\left.\partial_z\Phi(x,z)\right|_{z=0}=\left.\Phi(x,z)\right|_{z=\pi R}=0.
\end{equation}
We find
\begin{eqnarray} \label{eq:eo}
\nonumber
\Pi^{+-}_{S^1/Z_2}(q^2;R) &=& \Pi_{S^1}(q^2;R) - \Pi_{S^1}(q^2;R/2)\\
 &=& - {1\over 8\pi^2} \int_0^1 dx x \sqrt{1-x^2}\ln\left[2\cosh\left({\pi R\over 2}\sqrt{{q^2 x^2/4}+m^2}\right)\right],
\end{eqnarray}
which is finite.  Despite this, the boundary coupling constants receive logarithmically divergent radiative corrections.  The reason why such divergences do not appear in Eq.~(\ref{eq:eo}) is that from the 5D point of view, the zero mode correlator is a non-local quantity.  It can be obtained from a 5D effective action (with the $z$ dependence included) by integrating over the compact space.  In the process of doing this integration, the UV divergences of one boundary combine with those of the other to give the total UV cutoff dependence of the zero mode Green's function.  Because of its short distance nature, the coefficient the UV logarithm at $z=0$, where $\Phi$ satisfies Neumann boundary conditions, has the same value as the corresponding UV logarithm in the $(+,+)$ case, Eq.~(\ref{eq:ee}).  Using the same type of arguments described in~\cite{mark}, it can be shown that the UV logarithm at $z=\pi R$, where the field satisfies Dirichlet conditions, has the opposite sign.  This leads to the cancellation of divergences in Eq.~(\ref{eq:eo}), as well as the RG equations
\begin{equation}
\mu{d\over d\mu}\lambda^{+-}_0= - \mu {d\over d\mu}\lambda^{+-}_R=-{1\over 96\pi^2}.
\end{equation}

For a $(-,-)$  scalar,
\begin{equation}
\left.\Phi(x,z)\right|_{z=0}=\left.\Phi(x,z)\right|_{z=\pi R}=0,
\end{equation}
we have
\begin{eqnarray}
\nonumber
\Pi^{--}_{S^1/Z_2}(q^2;R) &=& {1\over 48\pi^2{\bar\epsilon}} + {1\over 32\pi^2}\int_0^1 dx x \sqrt{1-x^2}\ln\left[{{q^2 x^2/4} + m^2\over \mu^2}\right]\\
 & & {}+ {1\over 2}\Pi_{S^1}(q^2;R), 
\end{eqnarray}
and thus
\begin{equation}
\mu{d\over d\mu}\lambda^{--}_{0,R} = {1\over 96\pi^2}.
\end{equation}

The low energy predictions for the case of softly broken bulk symmetry follows closely the discussion of the previous section.  The only novelty here is the explicit $\mu$ dependence, which must be compensated by $\mu$ dependence of the boundary couplings.  However, if the symmetry breaking is by soft terms, in a minimal subtraction scheme the logarithmic dependence on $\mu$ can be absorbed into the definitions of the boundary couplings which are universal.  Subtleties arise when symmetry breaking occurs through boundary conditions.  For instance, suppose we break the $Z_2$ symmetry of our $U(1)\times U(1)$ model by assigning $(+,+)$ boundary conditions to $\Phi_1$ and $(-,-)$ to $\Phi_2$.  In $\overline{MS}$, for $\sqrt{q^2} R\ll 1$,
\begin{equation} 
{1\over g_1^2(q^2)} \simeq{\pi R\over g_5^2} + {\lambda_1}_0(\mu) + {\lambda_1}_R(\mu) +{1\over 48\pi^2}\ln(2\pi R\mu) -{1\over 24\pi^2}\left[\ln\left(2\pi R\sqrt{q^2}\right)-{4\over 3}\right],
\end{equation}
where ${\lambda_1}_{0,R}(\mu)$ are the running boundary couplings associated with the gauge group $U(1)_2$.  In the same limit, 
\begin{equation}
{1\over g_2^2(q^2)} \simeq {\pi R\over g_5^2} + {\lambda_2}_0(\mu) +{\lambda_2}_R(\mu) - {1\over 48\pi^2}\ln(2\pi R\mu) +\mbox{analytic in $q^2 R^2$}.
\end{equation}
Notice that the non-analytic dependence on $q^2$ has dropped out of this equation.  Since the zero mode of $\Phi_2$ has been removed by the boundary conditions, this result is simply a manifestation of decoupling in this scheme.  Because $\Phi_1$ and $\Phi_2$ satisfy different boundary conditions at the orbifold fixed planes, $Z_2$ symmetry is explicitly broken on those points so that in general we cannot expect equality of the boundary couplings.  Consequently, it seems impossible to make a prediction for the quantity
\begin{equation} 
\label{eq:bdpred}
{1\over g_1^2(q^2)} -{1\over g_2^2(q^2)} \simeq \mbox{bd. terms} + {1\over 24\pi^2}\ln(2\pi R\mu) -{1\over 24\pi^2}\left[\ln\left(2\pi R\sqrt{q^2}\right)-{4\over 3}\right]
\end{equation}
due to the unknown values of the boundary couplings.  To make progress, additional assumptions about their magnitude relative to the calculable quantum corrections are needed~\cite{flat}.  Because the theory becomes strongly coupled at a scale $\mu_s\sim 512\pi/g_5^2$ (at which point the quantum corrections to the effective couplings are of the same order as the tree-level result; see for instance
Eq.~(\ref{eq:pow})), it is customary to make the assumption
\begin{equation}
\lambda_0(\mu_s)\sim\lambda_R(\mu_s)\sim {1\over 96\pi^2}.
\end{equation}
Then according to this naive dimensional analysis (NDA) estimate, by choosing $\mu=\mu_s$ in Eq.~(\ref{eq:bdpred}) we may neglect with some temerity the tree-level boundary couplings relative to the calculable logarithmic correction $\ln(2\pi R\mu_s)\sim\ln(1024\pi/g_4^2)$ to obtain a prediction for the difference of the zero mode couplings at low energy
\begin{equation}
{1\over g_1^2(q^2)}-{1\over g_2^2(q^2)} \simeq {1\over 24\pi^2}\ln(1024\pi/g_4^2) -{1\over 24\pi^2}\ln\left(2\pi R\sqrt{q^2}\right) +{\cal O}(1/96\pi^2),
\end{equation}
where $1/g_4^2=\pi R/g_5^2\sim 1$ is the low energy coupling at tree level.  We note that in the case where symmetries are broken by boundary conditions, bulk quantum effects are universal and irrelevant as far unification is concerned.  This means that the extreme UV sensitivity that plagues the predictions in the type of models considered in the previous section is absent here (the UV scale enters only logarithmically), leading to robust results for low energy physics.

\section{Coupling Constant Evolution in AdS Models}
\label{sec:ads}

We now determine how the one-loop effects calculated in the previous section are modified by the presence of a background $\mbox{AdS}_5$ geometry.  First we compute the one-loop correction to effective low energy (zero mode) coupling constants in the AdS background, which for the remainder of this paper we will take to be Euclidean.  We will show that the UV structure (for instance the power counting of ultraviolet divergences) of the gauge field zero mode correlators is identical to what we found in flat space in the last section.  In general, this must be the case, since the physics in the UV comes from distances shorter than the curvature scale, and is therefore not sensitive to it. In particular we will see that the logarithmic divergences of our AdS models are the same as those in flat space, leading to the same RG flows.  For instance, like in the previous section, only the coefficients of  brane localized field strength operators are logarithmically renormalized, while the bulk coupling of the AdS gauge theory receives linearly divergent loop corrections and therefore does not run.

While the UV structure is the same, in curved spaces there can be new finite loop effects not present in flat space field theories.  In Section~\ref{sec:adszm}, we discuss the nature of curvature corrections to the low energy gauge couplings.  Whereas in flat space gauge field zero mode amplitudes are linearly dependent on bulk mass scales larger than the compactification scale, in AdS scalar field  masses smaller than the curvature enter only logarithmically in the one-loop correction to the gauge couplings.  (Masses larger than the scale $k$, on the other hand, give rise to loop corrections that are identical to what we find in flat space).  

Although the zero mode results are sufficient for deriving low energy predictions, the origin of the large logarithms that appear is not entirely clear from this analysis.  We provide a physical explanation of the features of zero mode amplitudes in Section~\ref{sec:adsplanck}.  There we will see that the logarithms arise from the running of Planck correlators down to a scale near the KK mass gap, followed by a matching procedure to the zero mode observables.  In Section~\ref{sec:adscft} we discuss how the large logarithms may be understood in terms of the AdS/CFT correspondence.

\subsection{Zero mode Observables}
\label{sec:adszm}

First consider the one-loop vacuum polarization due to a scalar field with $(+,+)$ boundary conditions, as in Eq.~(\ref{eq:ppbc}).  Using the results of the appendices, this is given by  
\begin{equation}
\label{eq:zm}
\Pi(q^2)  =  -{1\over 48\pi^2}\left[{1\over{\bar\epsilon}}+{1\over 2}\ln\left({\mu^2\over 4kT}\right)\right]+{1\over 16\pi^2} \int_0^1 dx  x \sqrt{1-x^2} \ln N_{++}\left({x\sqrt{q^2}\over 2}\right),
\end{equation}
where 
\begin{equation}
\label{eq:ppsum}
\ln N_{++}(p)  = -\ln\left|i_\nu(p/T) k_\nu(p/k)-i_\nu(p/k) k_\nu(p/T)\right|,
\end{equation}
with $i_\nu(z)$, $k_\nu(z)$ defined in Appendix~\ref{app:modesums}.   The $1/\epsilon$ pole here has the same origin as the one that appears in flat space in Eq.~(\ref{eq:ee}).  It simply indicates that the boundary gauge field couplings (such as those of Eq.~(\ref{eq:locff})) are logarithmically renormalized.  This follows again from dimensional analysis:  the boundary couplings are the only dimensionless couplings that arise in this theory.  Only these quantities get logarithmically divergent loop corrections.  The fact that such divergences are indeed confined to the boundaries of the space can also be understood by computing the 5D effective action for a background gauge field with dependence on the compact coordinate $z$.  Loop corrections to this quantity contain $1/\epsilon$ poles multiplying delta functions with support on the boundaries of the space, meaning the counterterms needed to renormalize the 5D action are those of Eq.~(\ref{eq:locff}).  Notice that in accord with our general discussion,  the coefficient of the poles, and therefore the RG flows, are identical to the one we found for a scalar with $(+,+)$ boundary conditions on the flat orbifold. 

While the RG equations are the same as in flat space, their application in the compact AdS background is slightly more  subtle.  Consider the zero mode coupling in the $\overline{MS}$ scheme
\begin{equation}
\label{eq:coup}
{1\over g^2(q^2)} = {R\over g_5^2} +\lambda_k(\mu) + \lambda_T(\mu) -{1\over 96\pi^2}\ln\left({\mu^2\over 4kT}\right) +{1\over 16\pi^2} \int_0^1 dx  x \sqrt{1-x^2} \ln N_{++}\left({x\sqrt{q^2}\over 2}\right),
\end{equation}
where $\lambda_{k,T}(\mu)$ are a set of running boundary gauge couplings, localized at the Planck and TeV branes respectively.  $R$ is the proper distance between the branes, which according to Eq.~(\ref{eq:metric}) is given by
\begin{equation}
R={1\over k}\ln\left({k\over T}\right).
\end{equation}
Although the effective coupling is $\mu$ independent, one would like to pick a value of $\mu$ in which all large logarithms have been resummed into the values of the boundary couplings.  One can actually avoid the issue of choosing a renormalization scale if we recognize that because the strong coupling scale for TeV brane correlators is of order $T$, we expect the boundary coupling $\lambda_T(\mu)$ to be given by its NDA estimate (of order $1/16\pi^2$) when evaluated at a renormalization scale $\mu\sim T$.  Likewise, the strong coupling scale for Planck localized Green's functions is $\mu\sim k$, so it is $\lambda_k(\mu\sim k)$ that we expect to be small on the basis of NDA.  Using the RG equations
\begin{equation}
\mu{d\over d\mu}\lambda_k(\mu)=\mu{d\over d\mu}\lambda_T(\mu)=-{1\over 96\pi^2},
\end{equation}
we can therefore relate the couplings evaluated at an arbitrary subtraction point $\mu$ to those at their NDA values.  In the process of doing so, the explicit logarithms of $\mu$ in Eq.~(\ref{eq:coup}) cancel with the logarithms that appear in the solution of the RG equation.  We are then left with 
\begin{equation}
{1\over g^2(q^2)} = {R\over g_5^2} +\lambda_k(2k) + \lambda_T(2T) + {1\over 16\pi^2} \int_0^1 dx  x \sqrt{1-x^2} \ln N_{++}\left({x\sqrt{q^2}\over 2}\right),
\end{equation}
where now all couplings are expected to be given by their natural values.  This is somewhat different from the flat space examples considered in the previous section, where the NDA scale was homogeneous across the compact direction.  Here, because of warping, we see that the one-loop corrections are cast in their simplest form when written in terms of the couplings renormalized at the scales associated with their location in the bulk spacetime\footnote{Equivalently, one could renormalize as in~\cite{GR} by performing the subtractions before taking the limit $D\rightarrow 4$.  While the interpretation of the boundary couplings differs in this scheme, the results for any physical quantity are not changed.}.

While the ultraviolet effects encoded in Eq.~(\ref{eq:zm}) are similar to those as in flat space, the low energy behavior for $\sqrt{q^2}\ll T$ differs.  We will separately consider the cases $m=0,$ $m<k$ and $m>k$.  For $m=0$ we have
\begin{equation}
\ln N_{++}(p) = -\ln{p^2\over kT}-\ln\left|I_1(p/T) K_1(p/k) - K_1(p/T) I_1(p/k)\right|,
\end{equation}
which for $p\ll T$ can be expanded as 
\begin{equation}
\ln N_{++}(p) \simeq -\ln{p^2\over T^2} + \mbox{terms analytic in $p^2/T^2$ or $p^2/k^2$}.
\end{equation}
Thus for $m=0$ and $\sqrt{q^2}\ll T$, we find
\begin{equation}
\label{eq:m0}
{1\over g^2(q^2)} = {R\over g_5^2}+\lambda_k(2k)+\lambda_T(2T) -{1\over 48\pi^2}\left[\ln{q^2\over T^2} -{8\over 3}\right] + \mbox{terms analytic in $q^2$}.
\end{equation}
Comparing to the one-loop vacuum polarization for a $(+,+)$ scalar field propagating on flat $R^4\times S^1/Z_2$, we see that the infrared logarithms of $q^2$ match those of Eq.~(\ref{eq:ee}) if we identify the flat space compactification scale $R$ with the parameter $T$ that appears here.   The fact that it is $T$, and not the brane separation $R$ that appears is a consequence of the  AdS curvature.  Given this fact, the infrared behavior is then identical to that of the flat 5D theory.  

In the massive case, we use the expansions for $z\ll 1$
\begin{eqnarray}
i_\nu(z) &\simeq& {2+\nu\over \Gamma(\nu+1)} \left({z\over 2}\right)^\nu\left[1+{\cal O}(z^2)+\cdots\right],\\
k_\nu(z) &\simeq& {\Gamma(\nu)\over 2}(2-\nu)\left({z\over 2}\right)^{-\nu}\left[1+{\cal O}(z^2)+\cdots\right],
\end{eqnarray}
where in both equations $\dots$ denotes terms higher order in $z^2$.  For $p\ll T$, we have
\begin{equation}
\ln N_{++}(p)\simeq -\ln\left[{\nu^2-4\over\nu}\right] - \nu\ln{k\over T} + \cdots,
\end{equation}
and thus, for $m\neq 0$
\begin{equation}
\label{eq:adsmpi}
{1\over g^2(q^2)} \simeq {R\over g_5^2}+ \lambda_k(k)+\lambda_T(T) - {1\over 48\pi^2}\left[\ln\left({m^2\over 2\nu k^2}\right) + \nu\ln\left({k\over T}\right)\right] + \mbox{analytic in $q^2$} ,
\end{equation}
where we have used $\nu^2=4+m^2/k^2$.  From this equation, we see that a massive field decouples from the low energy gauge force, in the sense that it contributes only through terms that are analytic in $q^2$ and can therefore be absorbed into the coefficients of local operators involving the gauge field strength and its derivatives.  In particular, heavy bulk fields give large $q^2$ independent contributions to $\Pi(q^2)$ which manifests themselves as a correction to the bare coupling $g_5$.  

If $m\gg k$, Eq.~(\ref{eq:adsmpi}) becomes
\begin{equation}
\label{eq:mbigk}
{1\over g^2(q^2)}\simeq {R\over g_5^2}+ \lambda_k(k)+\lambda_T(T) - {m R\over 48\pi^2},
\end{equation}
which is remarkably similar to the analogous flat space limit $\sqrt{q^2}\ll R$, $m\gg R$ (see Eq.~(\ref{eq:mgR}) and Eq.~(\ref{eq:ee})).  The fact that Eq.~(\ref{eq:mbigk}) is so sensitive to the bulk mass implies that it is equally sensitive to the contribution of tree-level operators to the low energy couplings.  Unification in AdS models at a scale larger than the curvature scale therefore encounters the same types of problems as power law coupling unification does in flat space.  Remarkably, however, for $m<k$, the low energy coupling is only logarithmically sensitive to bulk mass scales
\begin{equation}
\label{eq:mlessk}
{1\over g^2(q^2)}\simeq {R\over g_5^2}+ \lambda_k(k)+\lambda_T(T) - {1\over 24\pi^2}\ln{m\over T}.
\end{equation}
Generically, this formula will also receive corrections from insertions of higher dimension operators at tree-level.  These corrections manifest themselves either as terms that scale with the bulk mass as $m R$ or as terms that are analytic in $q^2/T^2$.    The former corrections originate from operators of the form
\begin{equation}
S\sim  {1\over\sqrt{M}}\int d^5 X\sqrt{G}  \Sigma F_{MN} F^{MN},
\end{equation}
where $\Sigma$ is some scalar field that develops a VEV of order $m/g_5$.   Using NDA we find that these operators lead to a breakdown of the calculation if the scalar mass $m$ is larger than $k$.  For $m<k$,  the large logarithms in Eq.~(\ref{eq:mlessk}) dominate the vacuum polarization and we expect the predictions for the  low energy couplings obtained by integrating out heavy particles to be reliable.

These results can be applied to the calculation of predictions for the low energy couplings in our $Z_2$ model, assuming that $\Phi_1$ acquires a symmetry breaking mass term $m_1<k$ (but much larger than the KK scale $T$), while $\Phi_2$ remains massless.  We also take both fields to have $(+,+)$ boundary conditions.  We then find from Eq.~(\ref{eq:m0}) and Eq.~(\ref{eq:mlessk})
\begin{equation}
\label{eq:adspred}
{1\over g^2_1(q^2)}-{1\over g^2_2(q^2)}\simeq {1\over 24\pi^2}\ln\left({\sqrt{q^2}\over m_1}\right),
\end{equation}
where as a consequence of the $Z_2$ symmetry, we have taken the tree-level $U(1)_{1,2}$ couplings equal.  This equations is just what we obtained in the 4D version of our $U(1)\times U(1)$ model.  In realistic GUT models based on the AdS hierarchy, the $X,Y$ bosons will play a role analogous to that of $\Phi_1$ in this model, while $\Phi_2$ plays the role of a Standard Model gauge field.  Therefore, if the 5D GUT symmetry is broken by the VEV of a bulk Higgs field, we expect to find that the relations among the Standard Model low energy gauge couplings implied by broken GUT symmetry are identical to what is found in the minimal Standard Model with an energy desert.  

We now describe how the above results are modified when the charged scalar field satisfies different boundary conditions.  For the case of $(-,+)$ boundary conditions, we find 
\begin{equation}
{1\over g^2(q^2)}  = {R\over g_5^2}+ \lambda_k(k/2) + \lambda_T(2T) +{1\over 16\pi^2}\int_0^1 dx x \sqrt{1-x^2}\ln N_{-+}\left({x\sqrt{q^2}\over 2}\right),
\end{equation}
where now
\begin{equation}
\ln N_{-+}(p) = -\ln\left|i_\nu(p/T) K_\nu(p/k) - I_\nu(p/k) k_\nu(p/T)\right|.
\end{equation}
As before, we have used the RG equations for the boundary couplings 
\begin{equation}
\mu {d\over d\mu} \lambda_k = - \mu {d\over d\mu} \lambda_T = {1\over 96\pi^2},
\end{equation}
to write the one-loop corrections in a form that does not have large UV logarithms.  For $\sqrt{q^2}\ll T$ this equation becomes
\begin{equation}
{1\over g^2(q^2)}  = {R\over g_5^2}+ \lambda_k(k/2) + \lambda_T(2T) -{1\over 48\pi^2}\left[\ln\left({2+\nu\over 2\nu}\right) +\nu\ln{k\over T}\right].
\end{equation}
In our toy GUT model, if $Z_2$ is broken by assigning $(-,+)$ boundary conditions to $\Phi_1$, while giving ordinary $(+,+)$ boundary conditions to $\Phi_2$ we have (both fields are kept massless)
\begin{equation}
\label{eq:qoverk}
{1\over g_1^2(q^2)}-{1\over g^2_2(q^2)}\simeq\mbox{bd. terms} +{1\over 24\pi^2}\ln\left({\sqrt{q^2}\over k}\right).
\end{equation}
where the unknown boundary corrections are expected to be small on the basis of NDA.  This result is somewhat surprising from the point of view of the KK expansion, where the modes of $\Phi_1$ and $\Phi_2$ are split at a scale of order $T$ (since $\Phi_1$ has no zero mode).  From this perspective, the $Z_2$ symmetry breaking scale is identified with the KK scale, and no large logarithms in the difference of couplings is expected.  The fact that a large logarithm appears in Eq.~(\ref{eq:qoverk}) is obvious, however, once one considers high energy observables.  

When we reverse the boundary conditions on the scalar from $(-,+)$ to $(+,-)$ we have instead
\begin{equation}
{1\over g^2(q^2)}  = {R\over g_5^2}+ \lambda_k(2k) + \lambda_T(T/2) +{1\over 16\pi^2}\int_0^1 dx x \sqrt{1-x^2}\ln N_{+-}\left({x\sqrt{q^2}\over 2}\right),
\end{equation}
with 
\begin{equation}
\ln N_{+-}(p) = - \ln\left|I_\nu(p/T) k_\nu(p/k) - i_\nu(p/k) K_\nu(p/T)\right|.
\end{equation}
We have also made use of the  RG equations 
\begin{equation}
\mu {d\over d\mu} \lambda_k = - \mu {d\over d\mu} \lambda_T = -{1\over 96\pi^2}.
\end{equation}
The low energy behavior depends on the value of the bulk mass.  For $m=0$ this is
\begin{equation}
{1\over g^2(q^2)}  = {R\over g_5^2}+ \lambda_k(2k) + \lambda_T(T/2) -{1\over 48\pi^2}\left[\ln\left({q^2\over T^2}\right)-{8\over 3}\right].
\end{equation}
Here, the logarithmic running below the scale $T$ is saturated at an exponentially small mass scale, of order $T^2/k$.  Inspection of the KK mass spectrum for a massless $(+,-)$ scalar field reveals that indeed there is an ``almost zero mode'' (that is, the wavefunction of the KK ground state is nearly flat) with such a small mass.  For non-zero mass this mode is lifted to the scale $T$ and we instead have
\begin{equation}
{1\over g^2(q^2)}= {R\over g_5^2}+ \lambda_k(2k) + \lambda_T(T/2) -{1\over 48\pi^2}\left[\ln\left({2-\nu\over 2\nu}\right) +\nu\ln{k\over T}\right].
\end{equation}
We see that again there is a logarithm of $m/T$ for bulk masses $m<k$.  In fact, both the massless and massive $(+,-)$ scalars give rise to coupling constant corrections that are identical to the corresponding $(+,+)$ examples worked out earlier.  So for instance, if symmetry breaking in our $Z_2$ model arises from a modification of TeV brane boundary conditions, no large logarithms will appear in the difference of low energy couplings.  In GUT models, we may want to assign $(+,-)$ boundary conditions to $X,Y$ bosons in order to suppress TeV brane proton decay.  This means that orbifold GUT breaking will not be sufficient to generate a reasonable prediction for the Standard Model gauge couplings measured at the weak scale.  It will be necessary to Higgs the GUT symmetry in the bulk as well.

\subsection{The Planck Brane Correlator}
\label{sec:adsplanck}

While it is straightforward to use a procedure like that of the previous section to compute predictions for low energy couplings in warped models, the physical origin of those results, particularly the large logarithms, is not completely clear from an analysis of KK modes alone.  Ideally, one would like to be able to understand how such logarithms arise from an EFT procedure in which one integrates out heavy bulk fields at a momentum scale of order their mass and then uses the RG to run the couplings down to low energies.  However, because of the power-law growth of zero mode correlators for energies larger than the KK gap, it is impossible to develop an EFT approach using such observables.  Furthermore, as we emphasized previously, in order to properly define high scale unification in a field theory context, it is necessary that there exist observables that are calculable at the GUT scale.  In the RSI scenario, Green's functions with external points lying on the Planck brane are insensitive to the effects of unknown UV physics as long as the typical external four-momenta are less than the scale $k$.  Consequently, they can be used to define the notion of a high GUT scale.  They can also be used for the purpose of understanding the evolution of couplings in an EFT approach.  

We will now show how to compute the gauge field Planck brane two-point correlator for external four-momenta $T\ll\sqrt{q^2}\ll k$. Doing so will enable us to understand how the large logarithms encountered in the previous section arise as a result of the usual matching and running of couplings in an EFT framework .  We do this at one-loop in massless AdS scalar electrodynamics with an action which includes 
\begin{equation} 
\label{eq:act} 
S={1\over 4g_5^2}\int d^5 X\sqrt{G} F_{MN} F^{MN} +\int d^5 X\sqrt{G}|D_M\Phi|^2,
\end{equation}
as well as terms such as those of Eq.~(\ref{eq:locff}).  To do this we need the gauge boson propagator, which in $A_z=0$ gauge is given by (here $\eta_{\mu\nu}$ is the flat Euclidean metric)
\begin{equation}
D^{\mu\nu}_q(z,z') = \eta^{\mu\nu} D_q(z,z') + {q^\mu q^\nu\over q^2} H_q(z,z'),
\end{equation}
where for one point on the Planck brane 
\begin{equation}
\label{eq:plprop}
D_q(z,1/k) ={kz\over q} {K_1(qz) I_0(q/T)+K_0(q/T)I_0(qz)\over I_0(q/T)K_0(q/k)-I_0(q/k) K_0(q/T)}.
\end{equation}
We will not need the specific form of $H_q(z,z')$.  The quantity of interest is
\begin{equation}
\label{eq:planck} 
\int d^4 x e^{ip\cdot x}\langle A_\mu(x,1/k) A_\nu(0,1/k)\rangle \equiv  {g^2(q^2)\over q^2} \eta_{\mu\nu}+\cdots,
\end{equation}
where we have defined an effective running coupling $g(q^2)$ as measured by Planck brane observers (and dropped gauge dependent pieces).    At one-loop, there are two diagrams which contribute.  The usual vacuum polarization graph is (dropping the longitudinal part of the gauge-boson propagator, and raising/lowering indices with the flat metric $\eta_{\mu\nu}$)
\begin{equation} 
\label{eq:l1} 
L^{(1)}_{\mu\nu}=\int {d^D p\over (2\pi)^D} {dz\over (kz)^3} {dz'\over (kz')^3}
D_q(1/k,z) S_{q+p}(z,z') (2p+q)_\mu (2p+q)_\nu S_p(z,z'),
D_q(z',1/k)
\end{equation}
whereas the seagull term of scalar QED is in this case
\begin{equation} 
\label{eq:l2}
 L^{(2)}_{\mu\nu}= -2\eta_{\mu\nu}\int {d^D p\over (2\pi)^D} {dz\over (kz)^3} D_q(1/k,z) S_p(z,z) D_q(z,1/k).
\end{equation} 
In these expressions, $S_p(z,z')$ is the scalar propagator\footnote{Strictly speaking we should have analytically continued our propagators and vertices to $\mbox{AdS}_{D+1}$.  In not doing so we miss out on constant terms which are irrelevant for our purposes here.}. We can understand the dominant non-analytic momentum dependence of these quantities by the following arguments.  First consider the spectral representation for the propagator of a massless scalar with $(+,+)$ boundary conditions
\begin{equation}
\label{eq:prop}
S_p(z,z')=\sum_n {\psi_n(z)\psi_n(z')\over p^2+m_n^2}.
\end{equation}
While the $n=0$ mode has $m_0=0$ and
\begin{equation}
\psi_0(z)=\sqrt{2k}\left[1-\left({T\over k}\right)^2\right]^{-1/2}\simeq \sqrt{2k},
\end{equation}
the excited states with $m_n < k$ are peaked towards the $z=1/T$ boundary.  However, in the limit  $T\ll\sqrt{q^2}\ll k$
\begin{equation}
D_q(z,1/k)\sim {k\over q}\sqrt{\pi z\over 2 q} {1\over K_0(q/k)}e^{-qz}.
\end{equation}
We therefore expect that all terms involving the $n\neq 0$ modes in the loop integrals will give contributions that are suppressed\footnote{If the field running in the loop has spin, the dominance of the zero mode over the excited KK states may no longer apply. However, it is still true in general that the regions of the loop integral away from $z=1/k$ are suppressed for external momenta less than the curvature scale.  Thus to a good approximation we may replace the propagators in the loop by their values on the Planck brane.  Given this fact, it is possible to generalize the statements made here regarding scalar fields to more realistic situations involving spinor and vector fields.} by powers of $T/k$ relative to the terms involving only $n=0$.   To see this explicitly, we note that in order to calculate the one-loop graphs we need the integrals
 \begin{equation}
I^{(1)}_{nm}=\int_{1/k}^{1/T} {dz\over (kz)^2} K_1(qz)\psi_n(z)\psi_m(z),
\end{equation}
and
\begin{equation}
I^{(2)}_{n}=\int_{1/k}^{1/T} {dz\over (kz)} K_1(qz)^2\psi_n(z)^2.
\end{equation}
The ratios $I^{(1)}_{nm}/I^{(1)}_{00}$ and $I^{(2)}_{n}/I^{(1)}_{0}$ may be calculated simply by noticing that for small values of $z$, where the integrands have their support, the excited states approach the constant values
\begin{equation}
\psi_n(z\simeq 1/k)\rightarrow - {\pi m_n\over 2\sqrt{k}} Y_1(m_n/T).
\end{equation}
From this we see that ${I^{(1)}_{n0}/I^{(1)}_{00}}\sim\sqrt{m_n T}/k$ for $n\neq 0$ and for $n$ or $m$ not zero ${I^{(1)}_{nm}/I^{(1)}_{00}}\sim{\sqrt{m_n m_n} T/k^2}$ (where in both cases we have taken the mode masses $T\ll m_n <k$).  Similarly for $n\neq 0$, ${I^{(2)}_{n}/I^{(2)}_{0}}\sim m_n T/k$. 
On the other hand, 
\begin{equation}
I^{(1)}_{00}\simeq {\psi_0^2\over 2q},
\end{equation}
is independent of $T$.

Given these facts, it is easy to calculate the leading one-loop corrections to the Planck two-point function of the gauge field.  We work in a decoupling scheme, so that as in a usual EFT calculation, for external momenta $\sqrt{q^2}\ll k$, we need only consider  KK modes with masses less than $\sqrt{q^2}$.  That is, modes with masses heavier than $\sqrt{q^2}$ give rise only to trivial contributions that can be absorbed into local counterterms\footnote{In this case, the proper local counterterm is the coefficient of the Planck brane localized gauge field strength operator.}.  Despite the fact that the integrals $I^{(1)}_{nm},I^{(2)}_n$ for the massive modes are suppressed by powers of $T$ relative to the integrals involving only the zero mode, it is possible that the sum over the roughly $\sqrt{q^2}/T$ KK states that contribute to the loop amplitude give rise to a term that is independent of $T$.  One can show that the leading $T$-independent term of this sum is of order $q^2/k^2$ and thus negligible for $\sqrt{q^2}\ll k$.   We are thus effectively only left with the contribution form the zero mode.  In this approximation, $L^{(1,2)}_{\mu\nu}$ is given by (in dimensional regularization)
\begin{eqnarray}
\label{eq:zml1}
 \nn
 L^{(1)}_{\mu\nu}  &=& {\psi_0^4\over 4 q^4 K_0(q/k)^2}\int {d^D p\over (2\pi)^D} {(2p+q)_\mu (2p+q)_\nu\over q^2 (q+p)^2}= {k^2\over q^4} {1\over K_0(q/k)^2}\left(q_\mu q_\nu - q^2\eta_{\mu\nu}\right)\Pi_{4D}(q^2)
\end{eqnarray}
 Note that the zero mode contribution to the loop integral in this equation is exactly what one finds for the one-loop vacuum polarization calculation in a purely 4D scalar QED calculation (which we denote by $\Pi_{4D}(q^2)$).  We have also dropped the zero mode contribution to the seagull term, since it vanishes in dimensional regularization.


Including tree-level effects as well as the contribution from the Planck brane boundary gauge coupling (the contribution of  the TeV brane gauge coupling is highly suppressed in the limit $\sqrt{q^2}\gg T$) 
\begin{eqnarray}
\nn
\int d^4 x e^{iq\cdot x}\langle A_\mu(x,1/k) A_\nu(0,1/k)\rangle  &=&  {k\over q^2}{g_5^2\over K_0(q/k)}\eta_{\mu\nu}\left[1-{k\over K_0(q/k)}g_5^2\lambda_k \right.\\
 & & \left. {} -  {k\over K_0(q/k)} g_5^2 \Pi_{4D}(q^2)+{\cal O}(g_5^4)\right] +\cdots,
\end{eqnarray}
where we have again ignored terms that depend on the choice of gauge.  Resumming the above terms we find that the effective Planck brane coupling at one-loop is
\begin{equation}
\label{eq:planckg}
 {1\over g^2(q^2)}= {1\over g_5^2 k} K_0(q/k)+\lambda_k(\mu) - {1\over 48\pi^2}\ln\left({q^2\over \mu^2}\right).
\end{equation} 
In this equation, the first term is due to the tree-level bulk gauge coupling.  Because
\begin{equation}
 K_0(q/k)\simeq-\ln\left({q\over 2k}\right),
\end{equation}
it is customary to think of this term as giving rise to a tree-level running of the coupling $g_5$.  However, this uncalculable contribution to the running is completely universal and thus irrelevant as far as unification is concerned.  Note that beyond this tree-level effect, the non-analytic, prescription independent momentum dependence is identical to the result derived from the zero mode calculation.  Furthermore, Eq.~(\ref{eq:planckg}) implies that $\lambda_k(\mu)$ satisfies an RG equation
\begin{equation} \mu
{d\over d\mu}\lambda_k(\mu) = -{1\over 24\pi^2},
\end{equation}
which is different than what we found for its running in our computation of the KK zero mode gauge correlator.  This is to be expected, since calculating in a non-decoupling scheme leads to a beta function which includes spurious effects of massive particles which normally would not be included
in the context of an EFT calculation.  Of course, physical quantities are not sensitive to this apparent discrepancy, since in the non-decoupling scheme the low energy matrix element will contain large logarithms which exactly cancel the effects of the of the spurious contribution to the beta function.  Eq.~(\ref{eq:planckg}) is identical to what we found in~\cite{US} by using AdS/CFT duality as it applies to AdS backgrounds compactified by branes. 

Eq.~(\ref{eq:planckg}) is valid for $\sqrt{q^2}<k$.  As $\sqrt{q^2}$ approaches $k$, modes with masses $m_n\sim k$ are no longer decoupled from the one-loop sums.  Such modes are not suppressed near the Planck brane, so they give an order one contribution to the correlator (relative to the one-loop zero mode) for large momenta.  In fact, for $\sqrt{q^2}>k$, the correlator becomes insensitive to the curvature scale, so it must behave like a flat space 5D gauge correlator.  This leads one to conclude that for large energies, the behavior is a non-analytic power law of the form $\sqrt{q^2}/k$, reflecting the breakdown of the Planck observable at a scale which is (up to a loop factor) of order $k$. 

If the bulk scalar has a mass $m$, then we can understand  its decoupling from the point of view of the Planck correlator.  The massive scalar has a KK mode with mass roughly $m/\sqrt{2}$~\cite{pomarol,ads} that, unlike other modes with masses below the curvature scale, is unsuppressed near $z=1/k$.  Therefore, as the external momentum becomes larger than the mass of this mode, a logarithm of $\sqrt{q^2}$ which is not suppressed by powers of $T/k$, is induced.  As in the zero mass case, the logarithm has the same coefficient as in a 4D calculation.  For external momentum less than $m$, this special mode decouples and the running of the correlator freezes out.   Thus this mode properly accounts both for the decoupling of the bulk field when $\sqrt{q^2}<m$ and for the correct momentum dependence when $\sqrt{q^2}>m$.

It is now possible to understand from an EFT point of view the results of the previous section.  Suppose $\Phi_1$ has a symmetry breaking mass term $m_1<k$.  Then $\Phi_1$ decouples at scales less than its mass, and the only source of relative running between the $U(1)$ couplings is due to the one-loop contribution of the $\Phi_2$ zero mode.  Matching at the scale $q^2=m_1^2$, we find from Eq.~(\ref{eq:planckg}) 
\begin{equation}
\label{eq:matching}
{1\over g_1^2(q^2<m_1^2)} = {1\over g_5^2 k} K_0(q/k) + \lambda_1(\mu) - {1\over 24\pi^2}\ln\left({m_1\over\mu}\right),
\end{equation}
which implies that at any other momentum scale larger than $T$
\begin{equation}
\label{eq:plpred}
{1\over g_2^2(q^2)} = {1\over g_1^2(q^2)} -{1\over 24\pi^2}\ln\left({\sqrt{q^2}\over m_1}\right).
\end{equation}

We can use this result to derive low energy predictions by noting that the Planck brane correlator matches on smoothly onto the KK zero mode gauge field correlator at low energy.  This  can be seen explicitly from the fact that at $\sqrt{q^2}\ll T$, Eq.~(\ref{eq:plprop}) approaches the zero mode gauge boson propagator $1/q^2$.  While we cannot use our calculation of the Planck correlator to explicitly calculate the matching to the zero mode, we can still use Eq.~(\ref{eq:planckg}) to capture the leading logarithmic behavior.  This is because the threshold effects of the light KK modes, whose contribution is unsuppressed as the external momentum reaches the scale $T$,  cannot induce large logarithms (since matching only involves logs of ratios of light masses).  We then expect the prediction of Eq.~(\ref{eq:plpred}) to coincide with the non-decoupling result of the previous section up to terms that are small relative to the large logarithm of $\sqrt{q^2}/m_1$.  Indeed, this is exactly what happens.

Finally, one can also use the Planck brane Green's functions to understand what happens when one of the scalars in our $Z_2$ model has either $(-,+)$ or $(+,-)$ boundary conditions.  In the $(-,+)$ case, there is no KK mode that has a strong overlap with the external gauge boson propagators.  The only logarithms of $q^2$ then come from the scalar field satisfying $(+,+)$ boundary conditions.  Because the $Z_2$ symmetry is explicitly broken on the Planck brane, there is no analog of Eq.~(\ref{eq:matching}).  Instead, we evaluate Eq.~(\ref{eq:planckg}) for the $U(1)_2$ gauge group at a renormalization scale $\mu\sim k$ where the boundary couplings have values that are small according to NDA.  This accounts for the logarithm of $\sqrt{q^2}/k$ that we found in the non-decoupling result of the previous section.  When the scalar is $(+,-)$, there is no zero mode, but recall that there is a near zero-mode, with a light mass of order $T^2/k$.  This mode is nearly flat, and is unsuppressed near $z=1/k$, so it gives rise to logarithmic running of the $U(1)_1$ coupling, again with the same coefficient as in 4D.  Up to small corrections due to the non-universal TeV brane coupling, the running of the two couplings is therefore identical, which explains why no large logarithms appeared in the low energy predictions of Section~\ref{sec:adszm}.

\subsection{CFT Interpretation}
\label{sec:adscft}

The results derived in the previous section can also be obtained via the AdS/CFT correspondence as it applies to the RS scenario.  As we discussed previously~\cite{US}, the Planck brane one-loop two-point correlator  is identical to the Green's function of a four-dimensional gauge field $A_\mu(x)$ in the dual 4D theory
\begin{equation}
\label{eq:dual}
{\cal L}_{\mbox{\tiny{4D}}} = {\cal L}_{\mbox{\tiny{CFT}}}  +
{1\over 4 g^2} F_{\mu\nu} F^{\mu\nu}  +  A_\mu J_{\mbox{\tiny{CFT}}}^\mu + + |D_\mu\phi|^2 + c\left(\phi{\cal O}_4 +\mbox{h.c.}\right).
\end{equation}
Here $A_\mu$ weakly gauges a $U(1)$ global symmetry of an (unknown) CFT described by  ${\cal L}_{\mbox{\tiny{CFT}}}$.  The scalar field $\phi(x)$ corresponds to the zero mode of our charged bulk  (massless) scalar $\Phi$, and couples to the CFT through the dimension-four operator ${\cal O}_4$ with a coupling $c$ which is of order $1/k$ (we ignore couplings to gravity).  

The running of the gauge coupling $g$ in Eq.~(\ref{eq:dual}) is exactly the running of the Planck correlator computed in the previous section~\cite{US}.  The former quantity can be extracted from the vacuum polarization of the 4D gauge field, which from Eq.~(\ref{eq:dual}) is given by 
\begin{equation}
\label{eq:picft}
\Pi^{\mu\nu}(q) = \int d^4 x e^{i q\cdot x} \left\langle J^\mu_{\mbox{\tiny{CFT}}}(x) 
J^\nu_{\mbox{\tiny{CFT}}}(0)\right\rangle_{\mbox{\tiny{CFT}}} - {1\over 48\pi^2}\left(q^\mu q^\nu - q^2\eta^{\mu\nu}\right)\ln\left({q^2\over\mu^2}\right) + {\cal O}(|c|^2).
\end{equation}
The first term represents the renormalization of the coupling due to pure CFT effects.  It is fixed in terms of Ward identities and conformal invariance to be 
\begin{equation}
\label{eq:JJ}
\int d^4 x e^{i q\cdot x} \left\langle J^\mu_{\mbox{\tiny{CFT}}}(x) J^\nu_{\mbox{\tiny{CFT}}}(0)\right\rangle_{\mbox{\tiny{CFT}}}={1\over 2 g_5^2 k}\left(q^\mu q^\nu - q^2\eta^{\mu\nu}\right)\ln\left({q^2\over k^2}\right),
\end{equation}
where the coefficient is known in terms of the 5D parameters due to the fact that this term is equivalent to the tree-level Planck gauge propagator in the AdS description~\cite{APR}.  The second term  of Eq.~(\ref{eq:picft}) is simply the one-loop contribution of the scalar $\phi$, while the ${\cal O}(c^2)$ corrections denote corrections to the vacuum polarization due to insertions of the operator ${\cal O}_4$.  These are suppressed when the external momentum is smaller than a  scale of order $k$.   From Eq.~(\ref{eq:picft}) we find that the running coupling at order $g^2$ is given by 
\begin{equation}
\label{eq:cftg}
{1\over g^2(q^2)} = {1\over g^2(\mu)} -{1\over g_5^2 k}\ln\left({q\over k}\right) -{1\over 48\pi^2}\ln\left({q^2\over \mu^2}\right),
\end{equation}
which exactly matches what we found in the AdS calculation.  For more details of the computation in the dual field theory, see our previous paper~\cite{US}.

The leading pure CFT correction to the running coupling, Eq.~(\ref{eq:JJ}),  is universal and therefore irrelevant for unification (indeed in 5D it represents the tree-level gauge propagator).  One may worry that there are subleading non-universal corrections to Eq.~(\ref{eq:JJ})  
which are as large as the one-loop contribution of the scalar $\phi$.  Computing them would be impossible without knowledge of the precise nature of the CFT which is  dual to our AdS theory. 
We note that this is not a problem, however.  First of all, we explicitly did the 5D calculation and found exactly Eq.~(\ref{eq:cftg}).  Furthermore, on the CFT side, we know exactly how non-universal effects are encoded in the dual 4D description, namely through the insertions of the dimension four operator (in the massless case)  ${\cal O}_4$.  This matches the fact that contribution of the KK modes (which in the dual picture are CFT bound states \cite{APR}) was found to be exponentially suppressed.

Now let us interpret  our results in the context of Eq.~(\ref{eq:dual}) for both mass term and boundary condition symmetry breaking in the toy $Z_2$ model.  Breaking $Z_2$ by an explicit bulk mass term $m$ for, say, $\Phi_1$ modifies the dynamics the corresponding scalar field in the 4D dual description.  In particular, the dual 4D scalar is a propagating degree of freedom only for energies scales above $m$.  This can be seen by noting that the massive bulk scalar in the AdS description is dual to the source of an operator in the CFT whose conformal dimension is given by $2+\sqrt{4+m^2/k^2}$.  It can be shown that if this operator is nearly marginal (i.e, $m\ll k$), then for energy scales above the mass $m$, quantum corrections in the CFT induce a kinetic term for this source, in which case it is promoted to a dynamical field.  Thus for $q^2>m^2$, both $U(1)$ couplings run equally, while for $q^2<m^2$ only the $U(1)_2$ coupling runs, giving rise  to a splitting of order $\ln(\sqrt{q^2}/m)$ between the low energy couplings, in agreement with our AdS results.

The case of orbifold symmetry breaking also has a simple CFT interpretation.  Consider first the case of $(+,-)$ boundary condition.  The fact that in 5D, $Z_2$ symmetry breaking is localized on the TeV brane is equivalent to the statement that in the dual theory the $Z_2$ symmetry is broken by the IR dynamics of the CFT.   Then it is clear that the relative running of the couplings will not induce large logarithms.  In the opposite $(-,+)$ case, the symmetry is explicitly broken on the Planck brane, which corresponds to high scale breaking in the dual picture.  This accounts for the large logarithm we found in the 5D calculation.

\section{Conclusion}

In this paper we have calculated the low energy couplings of gauge theories in AdS backgrounds by several distinct methods.  First, we straightforwardly computed in a non-decoupling scheme, finding, for scalar bulk masses less than the curvature scale, logarithmic sensitivity to masses of particles in the loops.  We  gave a natural interpretation of these results  in terms of an EFT calculation, by running Planck localized gauge field correlators to low energies and then matching to the zero mode quantities.  We have also extended the results of our previous paper to include a discussion of the case of symmetry breaking by boundary conditions. We also showed how to interpret both the mass and boundary condition breaking scenarios using AdS/CFT.  While here we only considered scalar charged fields, we expect similar results to arise in more realistic settings.  

Finally, one may wonder how the running of the gauge couplings is modified when the background spacetime deviates from $\mbox{AdS}_5$.  Although a full answer to this question is beyond the scope of this paper, our analysis can be easily generalized to consider what happens in more general warped backgrounds with Poincare symmetric 4D slices.  Briefly, we expect to find large logarithmic corrections to low energy couplings in any spacetime geometry which exhibits a pattern of  KK wavefunctions like AdS, in which 4D zero modes are delocalized, while excited KK states are localized away from the Planck brane.  Since the large logarithms are insensitive to the detailed structure to the IR (TeV) region of the space, it is likely that any space that is close to AdS in the UV (near the Planck brane), but arbitrary in the IR (near the TeV brane) will give rise to similar patterns for the running at one-loop.  More work is necesary to determine full range of possibilities for the running of the couplings in general warped backgrounds, however.

\section{Acknowledgments}

W.G. is supported in part by the DOE contract DE-AC03-76SF00098 and by the NSF grant PHY-0098840 and acknowledges the hospitality of the Aspen Center for Physics where some of this work was carried out.  The work of I.R. is supported in part by the DOE contracts  DOE-ER-40682-143 and DEAC02-6CH03000.  We thank the organizers of the Santa Fe Workshop on Extra Dimensions and Beyond, where this work was completed.

{\bf Note added:}  While this paper was being finished, ref.~\cite{cct} appeared which has some overlap with these results.

\appendix

\section{One-loop Vacuum Polarization for Compactified Theories}
\label{app:seff}

 A convenient way of obtaining the zero mode correlator is to calculate the corrections to the zero mode gauge field effective action due to integrating out bulk fields.  Let the higher dimensional gauge field obtain a classical background $A_\mu(x)$.  We first show how to compute the one-loop
vacuum polarization effects of a bulk scalar field to the effective action for $A_\mu(x)$.  The effects of fields in other representations of the Lorentz group can be obtained by a straightforward generalization of the results presented here. The one-loop bulk scalar contribution to the effective action can be obtained by summing the contribution of each 4D KK state.  In Euclidean signature, it is given by 
\begin{equation}
\label{eq:seff}
S_{eff}[A_\mu] = {1\over 4 g_4^2}\int
d^D x F_{\mu\nu} F^{\mu\nu} + \sum_n \mbox{tr}\ln\left[-D_\mu D^\mu + m_n^2\right],
\end{equation}
where $m_n$ are the KK masses, and $g_4$ is the effective 4D gauge
coupling.  For example, in a compactified 5D theory (possibly with
boundaries), this is given by
\begin{equation}
{1\over g_4^2} = {R\over g_5^2} + \sum_i \lambda_i,
\end{equation}
where $R$ is the volume of the compact manifold and $\lambda_i$ are the coefficients of a set of boundary localized gauge field kinetic operators.  Although we will only consider $U(1)$ gauge theories here, it is not difficult to include the one-loop corrections due to quantum fluctuations of a non-Abelian gauge field about its background value.

It is convenient to rewrite Eq.~(\ref{eq:seff}) as
\begin{equation}
S_{eff}[A_\mu] = {1\over 4 g_4^2}\int d^D x F_{\mu\nu} F^{\mu\nu} + {1\over 2}\int
 {d^D q\over (2\pi)^D} A_\mu(q) \Pi^{\mu\nu}(q^2) A_\nu(-q),
\end{equation}
with
\begin{equation}
\label{eq:pimunu}
\Pi^{\mu\nu}(q^2) = - \sum_n\int {d^D p\over (2\pi)^D}\left[{(2 p + q)^\mu (2p+q)^\nu\over (p^2+m_n^2)\left((p+q)^2 + m_n^2\right)}-{2\eta^{\mu\nu}\over p^2 + m_n^2}\right].
\end{equation}
Let
\begin{equation}
\label{eq:sums}
s(p) = \sum_n {1\over p^2 + m_n^2},
\end{equation}
then we can write
\begin{equation}
\sum_n {1\over (p^2 + m_n^2) (q^2 + m_n^2)} = {s(p)-s(q)\over q^2 - p^2}.
\end{equation}
Eq.~(\ref{eq:pimunu}) becomes
\begin{equation}
\Pi^{\mu\nu}(q^2) = (q^2\eta^{\mu\nu} - q^\mu q^\nu) \Pi(q^2),
\end{equation}
with
\begin{equation}
\label{eq:pi}
\Pi(q^2) = {2\over q^2}{1\over D-1}\int {d^D p
\over (2\pi)^D}\left[D-2 + {(q^2-4 p^2)\over q^2 + 2 q\cdot
p}\right] s(p).
\end{equation}
The integrals over momentum appearing in Eq.~(\ref{eq:pi}) can be
written in terms of hypergeometric functions.  Introduce
\begin{equation} 
I(z) = {\Omega_{D-1}\over (2\pi)^{D}}\int_0^\pi d\theta {\sin^{D-2}\theta\over 1 + 2z\cos\theta},
\end{equation} 
where $\Omega_D=2\pi^{D/2}/\Gamma(D/2)$ is the $D$-dimensional solid angle.  This integral can be obtained for instance from Eq.~(3.228.3) of~\cite{gr} by applying a chain of identities involving hypergeometric functions.  For $0<z<1/2$ the integral is
\begin{equation}
I(z) = {\Omega_D\over (2\pi)^D} F(1,1/2,D/2,4 z^2),
\end{equation}
while for $z<0$ or $z>1/2$ it can be written as
\begin{equation}
I(z) = {\Omega_D\over (2\pi)^D} {(D-2)\over 4 z^2} F(2-D/2,1,3/2,1/(4 z^2)).
\end{equation}
Defining
\begin{equation} 
\label{eq:N}
s(p) = -{1\over 2p}{d\over dp}\ln N(p),
\end{equation}
we may write
\begin{eqnarray}
\label{eq:blah}
\nn
\Pi(q^2) &=& {\Omega_D\over(2\pi)^D}{D-2\over 8}
\left({\sqrt{q^2}\over 2}\right)^{D-4}\left[\int_0^1 dx x^{D/2-2}
F(-1/2,1,D/2-1,x)\ln N\left({\sqrt{q^2 x}\over 2}\right)\right.\\
& & \left.{}+{D-4\over 3} \int_0^1 dx x^{1-D/2}
F(3-D/2,1,5/2,x)\ln N\left({\sqrt{q^2}\over 2\sqrt{x}}\right)\right].
\end{eqnarray}
This formula simplifies somewhat in four dimensions.  When $D=4$,
 the first term is finite.  Using
\begin{equation}
F(-1/2,1,1,x)=\sqrt{1-x}
\end{equation}
we end up with
\begin{equation}
\int_0^1 dx x^{D/2-2} F(-1/2,1,D/2-1,x)\ln N\left({\sqrt{q^2 x}\over
2}\right)\rightarrow 2 \int_0^1 dx x \sqrt{1-x^2}\ln N\left({x
\sqrt{q^2}\over 2}\right).
\end{equation}
On the other hand, depending on the asymptotic behavior of the integrand near $x=0$, the second integral is potentially divergent in four dimensions.  We will regulate this divergence by working
in $D=4-\epsilon$ dimensions.  If the function $N(p)$ defined in Eq.~(\ref{eq:N}) behaves as
\begin{equation} 
\ln N(p)\rightarrow
\beta_1+\beta_2\ln{p\over M} +\cdots
\end{equation}
as $p\rightarrow\infty$ (for constants $\beta_1$, $\beta_2$ and $M$), we then have
\begin{equation}
{D-4\over 3}\int_0^1 dx x^{1-D/2} F(3-D/2,1,5/2,x)\ln N\left({\sqrt{q^2}\over 2\sqrt{x}}\right)= -{2\over 3}\left[{\beta_2\over\epsilon} + \beta_2\ln\left({\sqrt{q^2}\over 2 M}\right)  + \beta_1 +{\cal O}(\epsilon)\right],
\end{equation}
and therefore
\begin{eqnarray}
\label{eq:it}
\nn
\Pi(q^2) &=& -{1\over 48\pi^2}\left[{\beta_2\over\epsilon}+\beta_2\ln\left({\mu\over M}\right)
+\beta_1 +{1\over 2}\beta_2\left(-\gamma + \ln(4\pi)\right)\right]\\
         & & {} + {1\over 16\pi^2}\int_0^1 dx x \sqrt{1-x^2}\ln N\left({x \sqrt{q^2}\over 2}\right).
\end{eqnarray}
where $\mu$ is an arbitrary subtraction scale.  It arises from the expansion of the factor $\left(\sqrt{q^2}\right)^{D-4}$ that appears in Eq.~(\ref{eq:blah}):
\begin{equation}
\left(\sqrt{q^2}\right)^{-\epsilon}=1-\epsilon\ln\left({\sqrt{q^2}\over\mu}\right).
\end{equation}
The $1/\epsilon$ pole corresponds to a logarithmic divergence, and is canceled by a similar $\epsilon$ dependence in the bare couplings appearing in Eq.~(\ref{eq:seff}).  The explicit $\mu$ dependence implies that the tree-level couplings themselves acquire a dependence on $\mu$ in such a way that any physical quantity derived from Eq.~(\ref{eq:seff}) is independent of the specific choice of scale. 

So far the discussion has been completely general and applies to fields propagating on compactified manifolds with arbitrary curvature or dimensionality.  To be concrete, let us apply this to models in flat 5D Euclidean space compactified either on $S^1$ or the line interval $S^1/Z_2$.  Taking the gauge group to be $U(1)$ and the scalar to have bulk mass $m$ (and periodic boundary conditions) and unit $U(1)$ charge, we find
\begin{eqnarray}
\nn
s_{S^1}(p;R) &=& \sum_{n=-\infty}^\infty {1\over p^2 + m^2 + n^2/R^2}\\
           &=& {\pi R\over \sqrt{p^2+m^2}}\coth{\left[\pi R\sqrt{p^2+m^2}\right]},
\end{eqnarray}
and thus
\begin{equation}
\ln N_{S^1}(p;R) = -2\ln\left[2\sinh\left(\pi R\sqrt{p^2+m^2}\right)\right].
\end{equation}
It follows from this that on $S^1$, $\beta_1=\beta_2=0$ and there
are no $1/\epsilon$ poles.  As discussed in the text, this is
consistent with the fact that on $R^4\times S^1$, loop corrections
do not give rise to logarithmic divergences.  Then
\begin{equation}
\Pi_{S^1}(q^2;R) = -{1\over 8\pi^2}\int_0^1 dx x
\sqrt{1-x^2}\ln\left[2\sinh\left(\pi R\sqrt{q^2
x^2/4+m^2}\right)\right].
\end{equation}
Given this result, it is straightforward to derive the
one-loop vacuum polarization effects of bulk fields on $R^4\times S^1/Z_2$.
 For instance for a scalar field satisfying $(+,+)$ boundary conditions
(defined in the text) at the fixed points of $S^1/Z_2$\begin{eqnarray}
s^{++}_{S^1/Z_2}(p;R) &=& \sum_{n=0}^\infty {1\over p^2 + m^2 + n^2/R^2}\\
               &=& {1\over 2}{1\over p^2 + m^2} +{1\over 2} s_{S^1}(p;R),
\end{eqnarray}
and consequently
\begin{eqnarray}
\nn
\Pi^{++}_{S^1/Z_2}(q^2;R) &=& {1\over 48\pi^2}\left[{1\over\epsilon} -{\gamma\over 2}\right]
 -{1\over 32\pi^2} \int_0^1 dx x \sqrt{1-x^2} \ln\left[{x^2q^2/4 + m^2\over 4\pi \mu^2}\right]\\
& &  +{1\over 2}\Pi_{S^1}(q^2;R)\;\;(+,+).
\end{eqnarray}
If on the other hand we choose $(+,-)$ boundary conditions with
the scalar vanishing on the boundary at $z=\pi R$ (but with
Neumann boundary conditions at $z=0$)  then
\begin{eqnarray}
\nn
s^{+-}_{S^1/Z_2}(p;R) &=& \sum_{n=0}^\infty {1\over p^2 + m^2 + (2n+1)^2/R^2}\\
\nn
               &=&  s^{++}_{S^1/Z_2}(p;R) - \sum_{n=0}^{\infty} {1\over p^2 + m^2 + (2n)^2/R^2}\\
               &=&  s^{++}_{S^1/Z_2}(p;R) - s^{++}_{S^1/Z_2}(p;R/2).
\end{eqnarray}
leading to the result in the quoted in the text.  Eq.~(\ref{eq:it}) can also be applied directly to our AdS examples, provided that the sums Eq.~(\ref{eq:sums}) over KK masses can be calculated.  We
present a method for doing this in the next appendix. 

\section{Mode Sums in AdS}
\label{app:modesums}

In Appendix~\ref{app:seff} we showed how to express the one-loop effective action for zero mode gauge fields in terms of sums over the KK masses of the form
\begin{equation}
s(p)=\sum_n {1\over p^2+m_n^2}.
\end{equation}
Our results were independent of the specific form of the KK mass spectrum.  Here we develop some tricks for evaluating sums of this type when the masses $m_n$ are the KK masses of bulk fields in compactified AdS backgrounds.  We will do this for an even-even scalar in the background of Eq.~(\ref{eq:metric}).  In that case the masses $m_n$ satisfy the equation
\begin{equation}
\label{eq:roots}
{\cal N}(m_n) \equiv j_\nu(m_n/T) y_\nu(m_n/k) - y_\nu(m_n/T) j_\nu(m_n/k) =0,  
\end{equation}
where $j_\nu(z) = (2-\nu)J_\nu(z) + z J_{\nu-1}(z)$ with $\nu=\sqrt{4+m^2/k^2}$, and $y_\nu(z)$ is similarly defined with $Y_\nu(z)$ replacing $y_\nu(z)$.  For the moment we will assume that the bulk mass $m$ is non-zero.  We now write
\begin{equation}
\sum_n {1\over p^2 + m_n^2} =\int_C {dz\over 2\pi i} f(z;p),
\end{equation}
where
\begin{equation}
f(z;p)={1\over p^2 + z^2} {d\over dz}\ln {\cal N}(z),
\end{equation}
and $C$ is a closed contour that encloses all the solutions of Eq.~(\ref{eq:roots}) on the real axis but excludes the points $z=\pm i p$ (we take $p$ real).  We can deform this contour to a contour $C'$ given by the union of the lines $z= i t$ with $t$ taking values in 
\begin{equation}
(-\infty,-p-\epsilon)\cup (-p+\epsilon,p-\epsilon)\cup(p+\epsilon,\infty),
\end{equation} 
and the semicircles $z=-ip + \epsilon e^{i\theta}$ and $z=ip +\epsilon e^{i\theta}$, with $\theta\in[-\pi/2,\pi/2]$ (taking $\epsilon\rightarrow 0$).  $C'$ also contains a circular arc that connects the endpoints $z=\pm i\infty$.  However the integral of $f(z)$ along this part of the contour is negligible.  Since in going from $C$ to $C'$ no poles of $f(z)$ are crossed, we have
\begin{equation}
\label{eq:c'}
s(p) =\int_{C'} {dz\over 2\pi i} f(z;p) = \Pr\int_{-\infty}^{\infty} {dt\over 2\pi} f(it;p) - {1\over 2}\mbox{Res} f(-ip;p) - {1\over 2}\mbox{Res} f(ip;p).
\end{equation}
But note that for $t\geq 0$ 
\begin{equation}
{\cal N}(\pm it) = {2\over \pi} \left[i_\nu(t/T) k_\nu(t/k) - k_\nu(t/T) i_\nu(t/k)\right], 
\end{equation}
where now $i_\nu(z)=(2-\nu) I_\nu(z) + z I_{\nu-1}(z)$, and $k_\nu(z)=(2-\nu) I_\nu(z) + z K_{\nu-1}(z)$ ($I_\nu(z)$ and $K_\nu(z)$ are the modified Bessel functions).  Then $f(it,p)$ is an odd function of $t$ and the principal value integral in Eq.~(\ref{eq:c'}) vanishes identically.  Therefore
\begin{equation}
s(p)={1\over 2p} {d\over dp}\ln\left[i_\nu(p/T) k_\nu(p/k) - k_\nu(p/T) i_\nu(p/k)\right], 
\end{equation}
and in the notation of the text
\begin{equation}
\ln N_{++}(p) = -\ln\left|i_\nu(p/T) k_\nu(p/k) - k_\nu(p/T) i_\nu(p/k)\right|,
\end{equation}
which together with the results of the previous appendix yields Eq.~(\ref{eq:ppsum}).  It is simple to generalize this method to fields with different spin or boundary conditions.  For instance, a scalar with odd boundary conditions on the Planck brane but even boundary conditions on the TeV brane ($(-,+)$) yields
\begin{equation}
\ln N_{-+}(p) = -\ln\left|i_\nu(p/T) K_\nu(p/k) - k_\nu(p/T) I_\nu(p/k)\right|.
\end{equation}
Similarly, the $(+,-)$ spectrum gives rise to 
\begin{equation}
\ln N_{+-}(p) = -\ln\left|I_\nu(p/T) k_\nu(p/k) - K_\nu(p/T) i_\nu(p/k)\right|.
\end{equation}


\end{document}